\documentclass[a4paper,11pt]{article}
\pdfoutput=1 % if your are submitting a pdflatex (i.e. if you have
\usepackage{jheppub} % for details on the use of the package, please
\usepackage[T1]{fontenc} % if needed

\def\be{\begin{equation}}
\def\ee{\end{equation}}
\def\bea{\begin{eqnarray}}
\def\eea{\end{eqnarray}}

\def\bma{\begin{pmatrix}}
\def\ema{\end{pmatrix}}

\def\bi{\begin{itemize}}
\def\ei{\end{itemize}}

\title{\boldmath Some Comments on  Dilaton Gravity.}

\author[a,1]{Enrique Alvarez,\note{Corresponding author.}}
\author[a]{Mario Herrero-Valea,}
\affiliation[a]{Departamento de F\'{\i}sica Te\'orica and Instituto de F\'{\i}sica Te\'orica, IFT-UAM/CSIC\\Universidad Aut\'onoma, 20849 Madrid, Spain}
\emailAdd{enrique.alvarez@uam.es}
\emailAdd{mario.herrero@estudiante.uam.es}
\abstract{
A Weyl invariant extension of Einstein gravity is studied. It simply consists in the group averaging of Einstein's action under Weyl transformations. Contradicting cherished beliefs, a conformal anomaly is found in the trace of the equations of motion if  diffeomorphism invariance is to be a symmetry of the quantum theory. This anomaly vanishes {\em on shell} which, according to general principles, means that there must exist a gauge in which it vanishes even {\em off shell}. It is however possible to keep Weyl invariance as a {\em bona fide} symmetry at the price of losing full diffeomorphism invariance. This is what happens in {\em unimodular gravity}, a closely related theory.}

\begin{document} 
{\flushright{IFT-UAM/CSIC-13-077, ~~FTUAM-13-17}}

\maketitle
\flushbottom

\section{Introduction}
Let us consider the following action principle (please cf. \cite{AH} for background and references).
\be\label{averaging}
S=\int d(vol)~\left(-{n-2\over 8(n-1)}~R~ \phi_g^2-{1\over 2}g^{\mu\nu}\nabla_\mu\phi_g\nabla_\nu \phi_g\right)
\ee
where we have represented the diffeomorphism invariant measure
\be
d(vol)\equiv \sqrt{|g|} d^n x
\ee
It  reduces to General Relativity (GR) in the gauge
\be
\phi_g=\sqrt{8(n-1)\over n-2} M_p^{n-2\over 2}
\ee
 and to {\em unimodular gravity} \footnote{Unimodular gravity is a speculative approach towards explaining why (the zero mode of) the vacuum energy seems to violate the equivalence principle (the {\em active cosmological constant problem}) is just to eliminate the direct coupling in the action between the potential energy and the gravitational field. This leads to consider unimodular theories, where  the metric tensor is constrained to be unimodular in the Einstein frame  $g_E\equiv \left|det~ g^E_{\mu\nu}\right|=1.$ 

The simplest nontrivial such unimodular gravitational action  reads
\bea
&&S_U\equiv -{1\over 16\pi G_n}\int d^n x ~R_E=- M_p^{n-2}\int d^n x~g^{1\over n} \left(~R+{(n-1)(n-2)\over 4 n^2} {g^{\mu\nu}\nabla_\mu g~\nabla_\nu g\over g^2}\right)\nonumber
\eea
This theory is Weyl  invariant  under
\be
\tilde{g}_{\mu\nu}=\Omega^2(x)~g_{\mu\nu}(x)
\ee
(the Einstein metric is inert under those) as well as under area preserving (transverse) diffeomorphisms, that is, those that enjoy unit jacobian, thereby preserving the Lebesgue measure. 
} in the gauge

\be\label{unimodulargauge}
\phi_g+2^{3\over 2} M_p^{n-2\over 2}\sqrt{n-1\over n-2} g^{-{n-2\over 4 n}}=0
\ee
  The field redefinition
\be
G_{\mu\nu}\equiv ~{1\over M_p^2}\left({n-2\over 8(n-1)}\right)^{2\over n-2}~\phi_g^{4\over n-2}~g_{\mu\nu}
\ee
reduces the theory to GR (modulo another boundary term)
\be
S=-M_p^{n-2}~\int \sqrt{G}~d^n x~R[G]
\ee
The Weyl symmetry 
\bea
&&\tilde{g}_{\mu\nu}=\Omega^2 g_{\mu\nu}\nonumber\\
&&\tilde{\phi_g}=\Omega^{2-n\over 2}~\phi_g
\eea
is then indeed a tautological symmetry to the extent that it leaves invariant the metric $G_{\mu\nu}$. This is non necessarily the case anymore when couplings to matter are considered, because we are going to assume that matter couples to $g_{\mu\nu}$ instead to $G_{\mu\nu}$.  Some interesting albeit speculative physical reasons as to why the metric $g_{\mu\nu}$ could be  the only one physically  observable have been advanced  by Gerard 't Hooft.
\par
Actually, in the present paper we shall confine ourselves to pure TWG in the absence of any matter. 
\par
In order to integrate over the gravitational fluctuations,  it  is much simpler to work with the singlet metric $G_{\mu\nu}$. There is an infinite factor coming from the functional integration over the gravitational scalar, which does not appear in the action. This infite factor disappears in all connected amplitudes. Although this point will be discussed at some length later on, let us be quite explicit here. We are {\em defining}
\bea
&& e^{i W\left[\bar{g}_{\mu\nu}, \bar{\phi}_g\right]}\equiv \int {\cal D} g_{\mu\nu}~{\cal D}\phi_g~e^{-i {1\over 2}\int d^4 x~\sqrt{-g}\left(\partial_\mu\phi_g g^{\mu\nu}~\partial_\nu\phi_g+{1\over 6}~R~\phi_g^2\right)}
\eea
through
\bea
&& e^{i W\left[\bar{G}_{\mu\nu}\left[\bar{g}_{\mu\nu},\bar{\phi}_g\right]\right]}\equiv \int {\cal D} G_{\mu\nu}~e^{{i\over 16\pi G} \int d^4 x R\left[G_{\mu\nu}\right]}
\eea
Actually there is in the best of cases a divergent proportionality factor, so that the equivalence is as best true for the connected piece, which we precisely denote the effective action, $W$. In the particular case of the Einstein-Hilbert term, the effective action is nothing but the well-known 't Hooft-Veltman \cite{'tHooftV} counterterm for pure gravity. This yields

\bea
&& S_{\infty}=\frac{1}{\pi^{2}(n-4)}\int d^4 x\sqrt{|G|}\left({149\over 2880}E_4[G]+{7\over 320} W_4[G]+{3\over 128}R[G]^2\right)
\eea
and $W_4$ is the square of Weyl's tensor, i.e.,
\be
W_4\equiv W_{\mu\nu\rho\sigma}W^{\mu\nu\rho\sigma}
\ee
The  Euler density (the quantity whose integral yields the Euler characteristic) is given by
\be
E_4\equiv R_{\mu\nu\rho\sigma}R^{\mu\nu\rho\sigma}-4 R_{\mu\nu}R^{\mu\nu}+R^2
\ee
Given the fact that the integral of the Weyl tensor squared is conformally invariant, we can naively put $G\rightarrow g$ on that term. If we keep the spacetime dimension at the generic value, the result is
\be
\int d(vol)~W_4\left[\Omega^2 g_{\mu\nu}\right]=~\int d(vol)~\Omega^{n-4}~W_4\left[g_{\mu\nu}\right]
\ee

This is due to the fact that the covariant Weyl tensor has conformal weight $-2$ in {\em any dimension}, whereas the volume element picks a factor $\Omega^n$. The same thing happens with the integral of the Euler density
 \be
\int d(vol)~E_4\left[\Omega^2 g_{\mu\nu}\right]=~\int d(vol)~\Omega^{n-4}~E_4\left[g_{\mu\nu}\right]
\ee

The term in $R^2$ is not conformal invariant in any dimension.
\par
This has the important consequence that the one loop expectation value of the {\em trace} of the equations of motion (this is the analogous to the energy-momentum tensor when gravity is dynamical) does not vanish
\be
\left\langle g^{\mu\nu}~{\delta S\over \delta g^{\mu\nu}}\right\rangle=2 \left.{\delta S_{eff}\over\delta\Omega}\right|_{\Omega=1}\neq 0
\ee

This is the analogous of the {\em conformal anomaly} and we shall dub it as such. 
In our case let us recall that the Weyl rescaling factor was promoted to a field as
\be
\Omega\sim ~\phi_g^{2\over n-2}
\ee
so that the correct Weyl transformation yields a factor of
\be
{1\over M_p^{n-4}}\left({n-2\over 4(n-1)}\right)^{n-4\over 2}\int d(vol)~\phi_g^{2(n-4)\over n-2}~W_4
\ee

If we were to keep the gravitational scalar with this dimensional dependent exponent in the action \cite{Englert} then the conformal anomaly would be absent.
\par
 The standard lore coming from dimensional regularization in flat space is however that the lagrangian must be fixed {\em before} regularization at a given spacetime dimension and only then define the bare lagrangian in dimensional regularization. The powers acting on the fields must in particular must be held fixed; this is the origin of the anomalous dimension of the coupling constants.
 \par
 Following these rules,  the conformal anomaly is given by
\be
\left\langle g^{\mu\nu}~{\delta S\over \delta g^{\mu\nu}}\right\rangle=2 \left.{\delta S_{eff}\over\delta\Omega}\right|_{\Omega=1}=\frac{1}{\pi^{2}}\int d^4 x\sqrt{|g|}~{7\over 320} W_4
\ee

\par
This was exactly what happened in  unimodular gravity  by using what amounts essentially to a different measure, so that the extension to the usual measure
\begin{equation}
\sqrt{|g|}d^4 x
\ee
to n-dimensions is not  diffeomorphism  invariant, but rather is taylored in such a way as to preserve Weyl invariance. For example
\begin{equation}
 \left(-g\right)^{2\over n}~W_4.
\ee
is point invariant in any dimension. 
This carries of course the punition of losing full diffeomorphism  invariance \footnote{ For the purposes of the present paper this is the same as {\em general covariance}, and we shall employ both terms as synonyms. It will be important in the sequel the difference with theories that are invariant under measure preserving diffeomorphisms only. For all practical purposes this means to treat the metric determinant $g\equiv det~g_{\alpha\beta}$ as a true scalar.
 }
; only measure preserving diffeomorphisms are maintained. 
\par
If we reject the presence of the dimension-dependent factor in the measure,
 \begin{equation}
 \sqrt{|g|}d^4 x\rightarrow |g|^{2\over n}~d^n x
 \ee
 arguing that $|g|$ is a composite field after all, and that its power must be fixed in any particular dimension once and for all, then {\em  even unimodular gravity would get a conformal anomaly}.

The total result for the divergent piece in four dimensions assuming diffeomorphism invariance is then
\bea
&& S_{\infty}=\frac{1}{\pi^{2}(n-4)}\int d(vol) \bigg\{{149\over 2880}E_4+{7\over 320} W_4+\left(R-6 {\nabla^2\phi_g\over\phi_g}\right)^2\bigg\}
\eea
The piece involving the gravitational scalar also yields a conformal anomaly, because the general formula
\begin{equation}
\left(\tilde{\nabla}^2-{n-2\over 4(n-1)}\tilde{R}\right)\left(\Omega^{-{n-2\over 2}}\phi\right)=\Omega^{-{n+2\over 2}}\left(\nabla^2-{n-2\over 4(n-1)} R\right)
\ee
implies that
\begin{equation}
\left(\tilde{R}- {4(n-1)\over n-2}{\tilde{\nabla}^2\tilde{\phi}_g\over \tilde{\phi}_g}\right)^2=\Omega^{-4}~\left(R- {4(n-1)\over n-2}{\nabla^2\phi_g\over \phi_g}\right)^2
\ee
which yields again a factor of $\Omega^{n-4}$ when combined with the  n-dimensional riemannian measure. The anomalous Ward identity of the four dimensional TWG then reads
\begin{equation}
\left\langle 0_+\left|-2 g^{\mu\nu}{\delta S_{TWG}\over \delta g^{\mu\nu}}-{n-2\over 2}\phi_g {\delta S_{TWG}\over \delta \phi_g}\right|0_-\right\rangle\equiv A_{TWG}=\frac{1}{\pi^{2}} \bigg\{{7\over 320} W_4+\left(R-6 {\nabla^2\phi_g\over\phi_g}\right)^2\bigg\}
\ee

This formula is the main result of this paper. The expression of the anomaly is manifestly  pointwise conformally invariant. 
It is interesting to compare this result with the cohomological analysis of Bonora, Cotta-Ramusino and Reina \cite{Bonora}. They admit only polynomial candidates for the cocycles.
The  cocycles which are not exact are
\bea
&& C_1\equiv W_4\nonumber\\
&&C_2\equiv E_4\nonumber\\
&&C_3\equiv \phi_g\nabla^2\phi_g-{1\over 6}R\phi_g^2\nonumber\\
&&C_4\equiv \phi_g^4
\eea

Our expression for the anomaly is clearly of the form
\begin{equation}
a C_1+ b {C_3\over C_4}
\ee
with a and  b constants.

\end{document}